%% file: main.tex
\documentclass[letterpaper,twocolumn,10pt]{article}

\input{packages}

\input{custom}

\usepackage{titling}

\begin{document}

\date{\vspace{-20pt}}

\title{\vspace{-10pt}\Large \bf From Models to Operators: Rethinking Autoscaling Granularity \\ for Large Generative Models\vspace{-10pt}}

\author{
{\rm Xingqi Cui}\\
Rice University
\and
{\rm Chieh-Jan Mike Liang}\\
Microsoft Research
\and
{\rm Jiarong Xing}\\
Rice University
\and
{\rm Haoran Qiu}\\
Microsoft Azure Research
} 

\maketitle

\input{000-abstract}
\input{001-intro}
\input{002-characterization}
\input{003-analysis}
\input{006-discussion}
\input{007-related}
\input{008-conclusion}

\begingroup
\raggedright
\bibliographystyle{plain}
\bibliography{references}
\endgroup

\end{document}

%% file: packages.tex
\usepackage{usenix-2020-09}  

\usepackage{tikz}
\usepackage{amsmath}

\usepackage{enumitem}

\usepackage[labelfont=bf]{caption}
\usepackage{subcaption}
\usepackage{cleveref}

\usepackage{amssymb,amsfonts}

\usepackage{algorithm}
\usepackage{algpseudocode}

\usepackage{booktabs}

\usepackage[normalem]{ulem}  

%% file: custom.tex
\usepackage{xspace}
\newcommand{\myparagraph}[1]{\vspace{\smallskipamount}\noindent\textbf{#1.\xspace}}
\newcommand{\myparagraphnodot}[1]{\vspace{\smallskipamount}\noindent\textbf{#1\xspace}}

\newcommand{\eg}{\emph{e.g.}\xspace}

\newcommand{\ie}{\emph{i.e.}\xspace}

\newcommand*{\rom}[1]{\uppercase\expandafter{\romannumeral #1\relax}}

\crefname{insight}{Insight}{Insight}
\newcounter{insight}
\newcommand{\boxinsight}[1]{%
  \refstepcounter{insight}%
  \vspace{\smallskipamount}%
  \noindent \simplebox{
    \textbf{\uline{Insight~\theinsight:}} 
    \textit{#1}
  }
}

\usepackage[most]{tcolorbox}
\tcbset{textmarker/.style={%
        enhanced,
        parbox=false,boxrule=0mm,boxsep=0mm,arc=3.5mm,
        outer arc=3.5mm,left=2mm,right=2mm,top=4pt,bottom=3pt,
        toptitle=1mm,bottomtitle=1mm,oversize}}

\newtcolorbox{simplenoteBox}{colback=white, colframe=black, boxrule=0.2mm, arc=1.5mm, auto outer arc, boxsep=0mm, left=2mm, right=2mm, top=1mm, bottom=1mm} 
\newtcolorbox{noteBox}{textmarker,
    colback=gray!8!white}

\newcommand{\simplebox}[1]{\begin{simplenoteBox} #1 \end{simplenoteBox}}

\usepackage{ifthen}
\newboolean{publicversion}
\setboolean{publicversion}{false}

\ifthenelse{\boolean{publicversion}}{
    \newcommand{\grumbler}[3]{}
    \newcommand{\ml}[1]{}
    \newcommand{\haoran}[1]{}
    \newcommand{\jx}[1]{}
    \newcommand{\todo}[1]{}
}
{
    \newcommand{\grumbler}[3]{\xspace\textcolor{#3}{\bf #1: #2}}
    \newcommand{\ml}[1]{\grumbler{ML}{#1}{violet}}
    \newcommand{\haoran}[1]{\grumbler{Haoran}{#1}{blue}}
    \newcommand{\jx}[1]{\grumbler{Jiarong}{#1}{red}}
    \newcommand{\todo}[1]{\textcolor{blue}{TODO: #1}}
}


%% file: 000-abstract.tex
\begin{abstract}
Serving large generative models such as LLMs and multimodal transformers requires balancing user-facing SLOs (e.g., time-to-first-token, time-between-tokens) with provider goals of efficiency and cost reduction.
Existing solutions rely on static provisioning or model-level autoscaling, both of which treat the model as a monolith.
This coarse-grained resource management leads to degraded performance or significant resource underutilization due to poor adaptability to dynamic inference traffic that is common online.

The root cause of this inefficiency lies in the internal structure of generative models: they are executed as graphs of interconnected \textit{operators}.
Through detailed characterization and systematic analysis, we find that operators are \emph{heterogeneous} in their compute and memory footprints and exhibit diverse sensitivity to workload and resource factors such as batch size, sequence length, and traffic rate. 
This heterogeneity suggests that the \emph{operator}, rather than the entire model, is the right granularity for scaling decisions.

We propose an \textbf{\textit{operator-level autoscaling}} framework, which allocates resources at finer (operator)-granularity, optimizing the scaling, batching, and placement based on individual operator profiles.
Evaluated on production-scale traces, our approach preserves SLOs with up to {40\%} fewer GPUs and {35\%} less energy, or under fixed resources achieves {1.6\texttimes{}} higher throughput with {5\%} less energy.
These results show that the operator, rather than the model, is fundamentally a more effective unit for scaling large generative workloads.
\end{abstract}

%% file: 001-intro.tex
\section{Introduction}

Serving online inference workloads of large generative models such as large language models (LLMs) and multimodal LLMs is both expensive and performance-sensitive.
End users expect fast responses from LLM services, often specified by service-level objectives (SLOs) on time-to-first-token (TTFT) and time-between-tokens (TBT).
Cloud providers, in contrast, focus on minimizing GPU cost, improving utilization, and reducing power or energy consumption.
Taming this tension is a core challenge in production model provisioning.

\myparagraph{Motivation}
Intuitively, LLM services can avoid SLO violations by statically provisioning for peak traffic profiles, \ie{}, sizing the number of model instance replicas to handle extreme cases.
However, from experiences at global cloud providers, LLM inference workloads exhibit a high degree of dynamics over time and hard-to-predict input/output sequence lengths; thus, reserved resources can only be utilized at {50\% and 39\%} when provisioning for the P95 demand of text and multimodal workloads, respectively~\cite{stojkovic2025dynamollm,qiu2025modserve,xiang2025aegaeon}.
This implies that static provisioning can incur a significant resource cost.

To handle this variability while meeting SLO requirements and achieving efficient GPU usage, the key enabler is SLO-targeted autoscaling of inference clusters.
At first glance, it appears that autoscaling can simply be implemented at a \textbf{\textit{model-level}} granularity, where the number of model replicas is adjusted to meet SLOs.
Despite its simplicity, model-level autoscaling fundamentally limits the ability to meet the two requirements above, as its coarse-grained strategy restricts fine-grained control and adaptability.
\uline{First}, model-level granularity treats the model as one monolithic scaling unit.
Generative models today are directed acyclic graphs (DAGs) of heterogeneous \textit{operators} such as attention, linear transformation, and normalization.
Effectively, scaling entire model replicas forces all operators to scale uniformly, even when only a small subset contributes to the bottleneck.
This \textit{over-provisioning} means that non-critical operators are also replicated to occupy precious GPU cycles, memory, and compute power that could otherwise be shared across workloads.

\uline{Second}, model-level autoscaling is slow: loading a full model onto additional GPUs incurs significant startup latency (\eg{}, loading a 70B model takes at least ten seconds on average even with state-of-the-art methods~\cite{fu2024serverlessllm}).
This delay makes it difficult to adapt quickly to traffic fluctuations that are common in LLM workloads~\cite{qiu2025modserve}, often resulting in SLO violations or costly over-allocation and thus poor utilization.

\begin{figure}[!t]
    \centering
    \includegraphics[width=\linewidth]{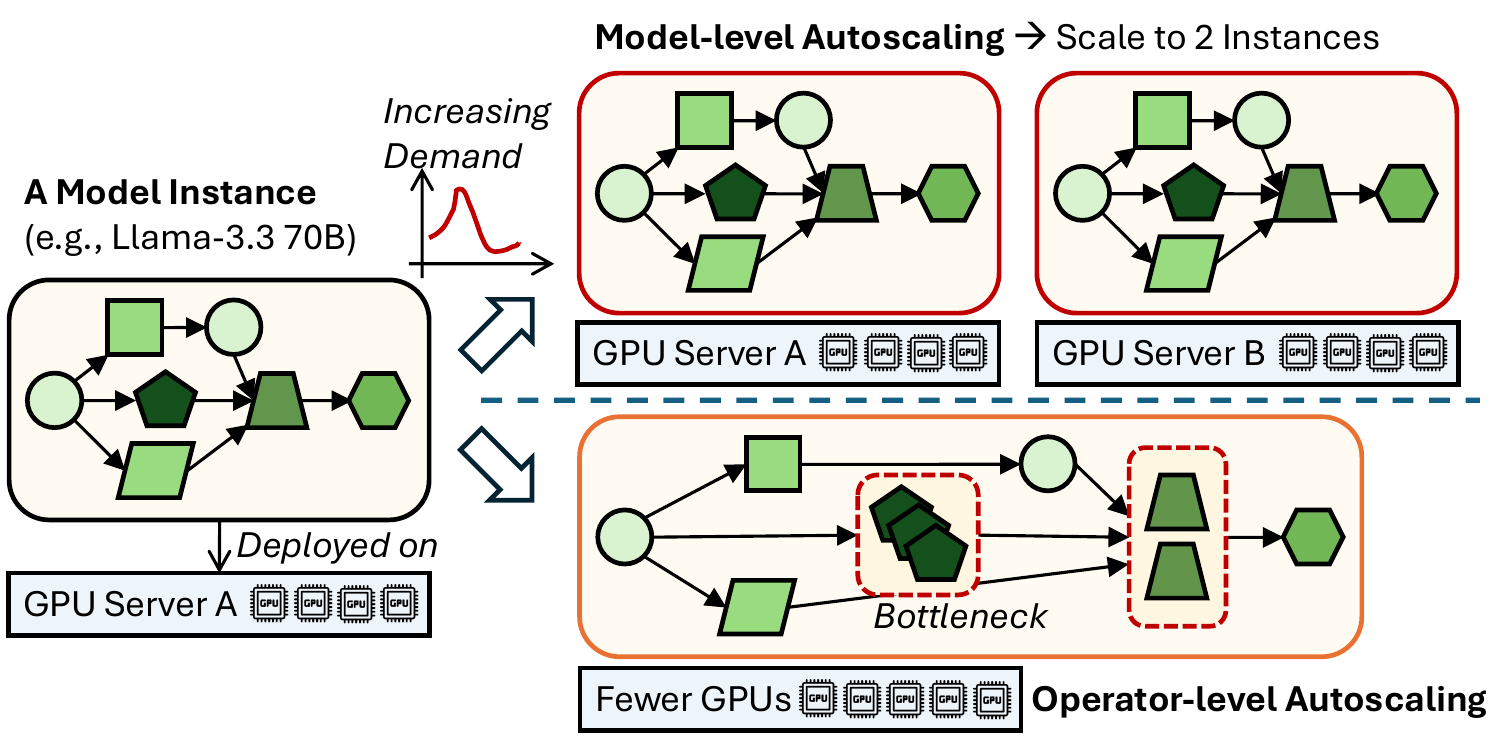}
    \caption{Operator-level vs. model-level autoscaling.\vspace{-10pt}}
    \label{fig:idea}
\end{figure}

\myparagraph{Our Work}
This paper explores a new opportunity for generative model provisioning: \textbf{\textit{operator-level autoscaling}} (as illustrated in \Cref{fig:idea}).
Rather than treating the model as one scaling unit, we propose to delve into operator-level granularity, for finer control of GPU resource allocation.
The goal is to independently scale the number of replicas for each operator, while trying to meet the model-level SLOs.
In addition, scaling at a finer granularity enables fast elasticity, reducing scaling latency from tens of seconds for model-level scaling (\eg{}, 10s for a 70B model~\cite{fu2024serverlessllm}) to sub-second response to handle rapid inference demand changes.

However, operator-level autoscaling raises two novel challenges.
\uline{First}, identifying which operators to scale is non-trivial, and na\"{i}vely targeting the slowest operator does not work well: the bottleneck shifts dynamically with workload changes (\eg{}, QPS, input lengths), and not all operators benefit equally from the same resource scaling.
For instance, attention operators are heavily compute-bound and sensitive to sequence length, especially under long context load, while linear operators dominate compute at short sequence lengths but scale less as the sequence length grows.
Operators like normalization are lightweight in compute, and benefit disproportionately from batching.
\uline{Second}, scaling operators with isolated device placement wastes resources, whereas colocating scaled operators improves utilization but risks interference across shared GPU resources (SMs, memory, interconnects), necessitating accurate contention modeling.

To address these challenges, we provide theoretical analysis with an SLO-oriented serving framework with operator-level provisioning that \uline{(1)} identifies scaling candidates that most effectively reduce end-to-end latency through offline profiling on both workload-aware performance sensitivity and resource elasticity analysis, and \uline{(2)} models colocation contention to guide interference-aware placement.
Architecturally, operator autoscaling operates across multiple planes of control:
First, the data plane captures the performance characteristics of each operator under diverse workload conditions (batch size, sequence length, query rate);
Then, the scaling plane models these profiles, and dynamically computes the scaling plan and configurations for all operators;
Finally, the scheduling plane takes the scaling plan, and jointly considers all operator replicas to compute the device assignment plan.
This enables more efficient utilization of GPU devices, better SLO preservation, and reduced energy consumption by rebalancing resources across operators to align supply with demand at a finer granularity than model-level autoscaling permits.

Across diverse model architectures (\Cref{tab:models}) on production-scale representative traces~\cite{stojkovic2025dynamollm,qin2024mooncake}, operator-level autoscaling achieves the same SLO preservation with up to {40\%} fewer GPUs and {35\%} lower energy consumption compared to model-level autoscaling.
With prefill-decode disaggregation, the resource savings achieved by operator-level autoscaling are most pronounced in the prefill phase (\ie{}, 2--3\texttimes{} higher than those of the decode phase), which highlights distinct computational profiles between two stages and suggests that inference clusters can gain substantial efficiency by enabling prefill–decode disaggregation.
These results demonstrate that each operator is a fundamentally more effective unit of scaling than the model as a monolith, enabling systems to balance end-user performance requirements with provider efficiency goals more precisely and promptly.

\myparagraph{Contributions}
We make the first step toward fine-grained resource management at the operator level for efficient generative model serving in cloud datacenters.
In summary, our main contributions include:
\begin{itemize}[leftmargin=*,nosep]
    \item Characterization of diverse operator compute–memory sensitivities under varying workload conditions.
    \item An operator-level autoscaling strategy that exploits operator heterogeneity for dynamic resource allocation.
    \item A contention-aware operator placement strategy that balances execution efficiency and cost.
    \item Extensive evaluation with production-scale traces, highlighting improvements in resource efficiency, throughput, and energy consumption across model architectures.
\end{itemize}

\input{tables/models}

%% file: tables/models.tex
\begin{table}[t]
\centering
\caption{Models in characterization study.
}
\label{tab:models}
\resizebox{\linewidth}{!}{%
\begin{tabular}{lccc}
\toprule
\textbf{Model} & \textbf{Model Size} & \textbf{Modality} & \textbf{Architecture} \\
\midrule
Qwen2-7B~\cite{qwen2-7b}        & 7B   & Text & Dense LLM \\
Qwen2-MoE~\cite{qwenmoe}     & 57B (14B)  & Text & MoE LLM \\
Llama3-8B~\cite{llama3}         & 8B   & Text & Dense LLM \\
Mixtral-8$\times$7B~\cite{mixtral} & 47B (13B) & Text & MoE LLM \\
Qwen2.5-VL-32B~\cite{qwenvl}        & 32B  & Visual & Encoder+LLM \\
\bottomrule
\end{tabular}}
\end{table}

%% file: 002-characterization.tex
\section{Background}
\label{sec:bg}

\input{figures/characterization/comp-sensitivity}

\subsection{GPU Execution Model for Inference}
\label{sec:bg:gpu}

Computation on GPU is expressed as \textit{kernels}, which run across many threads organized into \textit{blocks} and scheduled to run on \textit{Streaming Multiprocessors (SMs)} in parallel.
Each SM contains multiple SM cores---the fundamental compute units responsible for executing kernels.
This parallelism is essential for matrix and tensor operations that make up neural network layers in model inference.

For generative models like LLMs, inference consists of a sequence of \textit{operators} such as matrix multiplications, normalization, and attention.
These operators form a \textit{directed acyclic graph (DAG)} where edges represent tensor dependencies and nodes correspond to computation kernels.
Examples such as residual connections, layer normalizations, and attention branches create multiple dataflow paths within this DAG. However, as outputs depend only on previously computed tensors, the graph remains acyclic.
These operators are compiled into GPU kernels and run as part of a GPU \textit{stream}, which is an ordered sequence of kernels that can overlap with data transfers for efficiency~\cite{chen2018tvm}.
Model weights and intermediate activations reside in GPU memory, while temporary buffers handle transient memory needs during operations.
Multiple streams can run concurrently on GPUs to maximize hardware utilization with time or space sharing techniques like MPS~\cite{mps}, MIG~\cite{mig}, and CUDA Green Context~\cite{green-context}.

\subsection{Generative Model Inference Autoscaling}
\label{sec:bg:autoscaling}

A generative model inference cluster manages a pool of GPU servers to handle requests for models like LLMs, multimodal models, or diffusion models (\eg{}, for video generation).
These services back diverse applications such as interactive chat~\cite{chatgpt}, deep research jobs with relaxed deadlines~\cite{deepresearch}, or real-time audio streaming~\cite{realtime}.
They have varying SLOs and dynamic traffic patterns.
Two most common SLOs are on time-to-first-token (TTFT) and time-between-tokens (TBT).
This variability creates a fundamental tension: optimal autoscaling must allocate just enough server capacity to preserve user-centric SLOs (especially tail latencies) while minimizing provider-centric goals in cost and energy.



Today’s systems rely on \textit{\textbf{model-level autoscaling}} that combines \textit{horizontal autoscaling} (\ie{}, scaling in/out model instances) and \textit{vertical autoscaling} (\ie{}, scaling up/down model parallelism degrees).
Vertical autoscaling tunes intra-replica capacity via tensor or pipeline parallelism based on profiled memory footprint and latency-throughput tradeoffs~\cite{stojkovic2025dynamollm}.
Horizontal autoscaling typically combines demand forecasting with queueing-delay controllers and per-model replica scaling, often using signals like tokens-per-sec, queueing delays, and SLO violations~\cite{team2025aibrix,huang2024enova,patke2025hierarchical}.
However, short traffic bursts and rapidly varying context lengths frequently cause capacity misalignments and SLO regressions, as provisioning new model replicas is slow.
In addition, state-of-the-art autoscalers such as AIBrix~\cite{team2025aibrix}, DynamoLLM~\cite{stojkovic2025dynamollm}, vLLM Production Stack~\cite{production-stack}, and Chiron~\cite{patke2025hierarchical} fail to take advantage of operator-level heterogeneity for cost saving, i.e., not every operator is equally sensitive to workload changes.

In this paper, we systematically explore the benefits, opportunities, and challenges of \textbf{\textit{operator-level autoscaling}} and finer-grained resource management in modern generative model serving clusters.

\section{Operator Characterization and Insights}
\label{sec:bg:characterization}

We present the first systematic characterization of the performance and resource characteristics of each individual operator across large generative models.
Specifically, we study two dominant model architectures: dense LLMs and mixture-of-experts (MoE)-based architectures. We profile their compute, memory, input/output data volume, and queueing characteristics at the operator level, as well as the impact of GPU resource partitioning.
Characterizing these operators provides insights into their \textit{sensitivity} to different system factors such as sequence length, batch size, request arrival rate, and hardware allocation. For each operator, we define sensitivity as the normalized latency relative to a baseline configuration (\eg{}, batch size of 1 and the shortest sequence length).

\myparagraph{Experiment Setup}
We conduct experiments on vLLM across diverse models, including Qwen2-7B\cite{qwen2-7b}, Qwen2-MoE\cite{qwenmoe}, Llama3-8B\cite{llama3}, Mixtral-8x7B~\cite{mixtral}, and Qwen2.5-VL~\cite{qwenvl} (as listed in \Cref{tab:models}), on an Azure GPU server~\cite{a100azure} with 8 NVIDIA A100 GPUs.
For each model, we evaluate the inference runtime across multiple model configurations, varying prompt lengths, batch size, and tensor parallelism.
To characterize performance, we employ the CUDA time profiler to collect GPU kernel runtimes for each operator, based on vLLM's built-in layerwise-profile context manager to instrument the inference execution.
This setup captures fine-grained performance metrics for each kernel and operation, enabling a comprehensive analysis of runtime bottlenecks and operator-level summaries, including compute time, weight memory, activation memory, and inter-operator communication volume and data shape.
In addition, we leverage NVIDIA MPS~\cite{mps} to control the allocation of SM cores to individual operators (in the SM sensitivity experiment), and use NVIDIA DCGM~\cite{dcgm} to monitor SM utilization at runtime.

\input{figures/characterization/mem-sensitivity}
\input{figures/characterization/batch-size-sensitivity}

\myparagraph{Compute Characteristics}
We measure compute time as the actual GPU execution time per operator, recorded in microseconds ($\mu$s) using CUDA event profiling.
This metric isolates the pure computational cost of GPU kernels by excluding CPU overheads and memory-transfer latencies.
To understand how operator latency scales under different workloads, we evaluate its \textit{sensitivity} to key generative-model serving parameters (\ie{}, sequence length and batch size).
By tracking latency growth with increasing batch size or sequence length, we quantify each operator’s scaling behavior for more accurate performance modeling.

\Cref{fig:comp-sensitivity} shows that the compute sensitivity to sequence length is dominated by the attention operator.
In the prefill stage, self-attention exhibits quadratic time complexity with respect to sequence length $L$ (\ie{}, O($L^2 d$) with batch size $d$), since every token attends to all previous tokens.
By contrast, other operators—such as feed-forward layers, layer norms, and embedding lookups—scale linearly (\ie{}, O($L d$)) with sequence length and thus show far less increase in normalized latency.
Operators like softmax, fill, and sigmoid show nearly flat curves in compute sensitivity to sequence length in MoE models.
During decoding, we observe a similar trend across operators, but with smaller slopes, as the cached KV pairs reduce the computational cost of attention.
These findings highlight that prefill attention remains the key scaling challenge for generative model serving under long sequence lengths, regardless of model architecture.

\Cref{fig:batching-sensitivity} shows the compute sensitivity to batch sizes.
While most operators exhibit roughly linear scaling with batch size, there is still notable variation in slope across operators, which reflects differences in compute intensity.
Operators with heavier per-token arithmetic (\eg{}, large linear projections, fused MoE layers) scale closer to perfectly linear because their compute dominates memory overhead.
In contrast, lighter operators (\eg{}, layer norms, elementwise activations, small projections) show sub-linear scaling, as fixed kernel launch costs and memory-bound behavior become relatively more significant at larger batch sizes.
Thus, even though attention itself becomes linear with batch size, the variation between operators exposes their differing compute-to-memory ratios and can help identify which kernels are most sensitive to batching and which are bottlenecked elsewhere.
This reinforces the importance of per-operator profiling rather than assuming uniform scaling across the entire model.

\boxinsight{Operator compute sensitivity varies widely, with attention dominating across model architectures due to quadratic complexity. MoE and Encoder-LLM models exhibit more operators with flat scaling curves.}
\label{insight:comp-sensitive}

\myparagraph{Memory Characteristics}
Memory footprint is a key constraint for large-model serving, as GPUs must hold not only model parameters but also activations and the key–value (KV) cache during generation.
Therefore, memory profiling considers both weight memory and activation memory.
Weight memory corresponds to the static storage of model parameters, while activation memory refers to the intermediate tensors and KV cache generated during forward pass computation that depend on request sequence lengths.
By monitoring operator memory usage under varying sequence lengths, we quantify each operator’s memory sensitivity (similar to compute sensitivity) in terms of how its memory usage scales with workload dimensions to identify memory characteristics.

In transformer models, attention operators dominate memory growth due to their O($L^2$) scaling with sequence length, while most other operators grow roughly linearly.
However, with FlashAttention~\cite{dao2022flashattention}, the attention operator has linear memory complexity to sequence length due to I/O-aware optimizations.
Therefore, we observed that \texttt{act\_and\_mul} fused kernel, together with other linear kernels, have a similar growth trend compared to the attention operator across all models in \Cref{fig:mem-sensitivity}.
Lightweight operators like index‐select and activation kernels show flatter growth.

\input{figures/characterization/combined}

Combining compute and memory sensitivity, \Cref{fig:combined-comp-memory} shows that, for a layer of the Llama2-7B model, some operators are primarily memory-intensive (\eg{}, norm), while others are primarily compute-intensive (\eg{}, reshape and cache).
Certain operators, such as attention, are intensive in both dimensions.
This suggests that different scaling strategies are required depending on the operator’s resource profile.

\boxinsight{Memory sensitivity is more evenly distributed across operators compared to compute sensitivity, consistent across model architectures. The two dimensions are uncorrelated---operators may be intensive in both, one, or neither. Unlike compute, memory sensitivity is bounded by linear scaling with FlashAttention.}
\label{insight:mem-sensitive}

\input{figures/characterization/queueing-sensitivity}

\myparagraph{Queueing Characteristics}
Building on the per-operator compute sensitivities, we analyze how operators respond to increasing request load (RPS) using M/M/c queueing theory.
Each operator is modeled as a multi-replica queueing system, where the service rate is $\mu =$ 1/(op\_latency × num\_layers), derived from the measured GPU execution times, and the arrival rate is $\lambda=$ requests\_per\_second/batch\_size.
Using the Erlang-C formula, we estimate waiting times and determine the minimum number of replicas required to maintain system stability under varying RPS.

\Cref{fig:queueing-sensitivity} shows heterogeneous queueing sensitivities across operators that closely reflect their compute characteristics.
For attention operators, especially at longer sequence lengths, the number of replicas required grows sharply with increasing RPS, reflecting their high per-token computational cost.
In MoE models like Mixtral, the FusedMoE linear operator dominates compute at short sequence lengths, leading to a pronounced scaling of replicas even for relatively small sequences.
In contrast, lighter operators, such as layer norms and embeddings, exhibit moderate sensitivity to RPS.
Queueing delays increase non-linearly when replication is insufficient, so small reductions in replicas can cause disproportionately high waiting times.
These observations highlight that combining compute profiling and queueing modeling enables precise, operator-specific replication strategies by selectively replicating high-demanding operators (\eg{}, attention) while avoiding overprovisioning lightweight ones, thereby meeting end-to-end latency and throughput targets efficiently.

\boxinsight{Operators exhibit diverse queueing sensitivity with increasing load. Insufficient replication causes non-linear queueing delays, emphasizing the need for operator-specific replication strategies.}
\label{insight:queueing}

\input{figures/characterization/io-sensitivity}

\myparagraph{Dataflow Characteristics}
We analyze data flows by quantifying the communication payload between adjacent operators in the model graph.
Specifically, we perform transient memory profiling to capture the input-output scaling with sequence length and batch size for each operator, revealing a linear growth of data volume (as shown in \Cref{fig:io-sensitivity}).
Communication overhead is estimated from transfer latency, which scales proportionally with data volume.
Attention and linear operators exhibit near-constant per-request data volume, whereas attention and linear operators scale with sequence length.
Comparative analysis of compute versus NVLink transfer time shows transfer overhead reaching~20\% for certain operators (\eg{}, \texttt{SiLu Mul}) but remaining below 5\% for most.

\boxinsight{Data volume scales linearly or remains flat with sequence length across operators. Transfer overhead can reach~20\% of compute time, making transfer costs non-trivial when placing operators across devices.}
\label{insight:io}

\input{figures/characterization/sm-characterization}

\myparagraph{Sensitivity to SM Allocation}
Placing operators on GPUs, especially when multiple workloads share the same GPU, requires a precise understanding of how Streaming Multiprocessor (SM) allocation affects operator performance. This is similar to our compute characteristics study.
\Cref{fig:sm-characterization} illustrates how different operators respond to changes in SM allocation, which is controlled by the MPS (Multi-Process Service) percentage~\cite{mps}.
This analysis compares a long sequence length of 2K (prefill phase) and a short sequence length of 1 (decode phase).

For the prefill phase (2K sequence length), the top-left plot shows that as the MPS percentage increases, the normalized latency for all operators decreases significantly. We note that compute-intensive operators like Attention and MLP dominate latency, especially at lower MPS values.
This is because these operators saturate the SM utilization at limited resources, and therefore, a reduction in SM allocation directly increases their latency, explaining the steep performance curve.

In the decode phase (sequence length of 1), a different pattern emerges.
As shown in the top-right plot, the normalized latency for most operators remains low and relatively flat, with only a minor decrease as MPS increases. The bottom-right plot clarifies the reason: the SM utilization for these operators is low, and they do not saturate the available resources.
Therefore, reducing the MPS percentage has a minimal impact on their latency. This makes a lower MPS percentage suitable for short sequences, as it allows for better resource sharing without significantly compromising performance.

\boxinsight{Operator sensitivity to SM allocation varies widely across operators, sequence lengths, and prefill/decode phases, correlating with SM utilization patterns.}
\label{insight:sm}

%% file: figures/characterization/comp-sensitivity.tex
\begin{figure*}
    \centering
    \includegraphics[width=\textwidth]{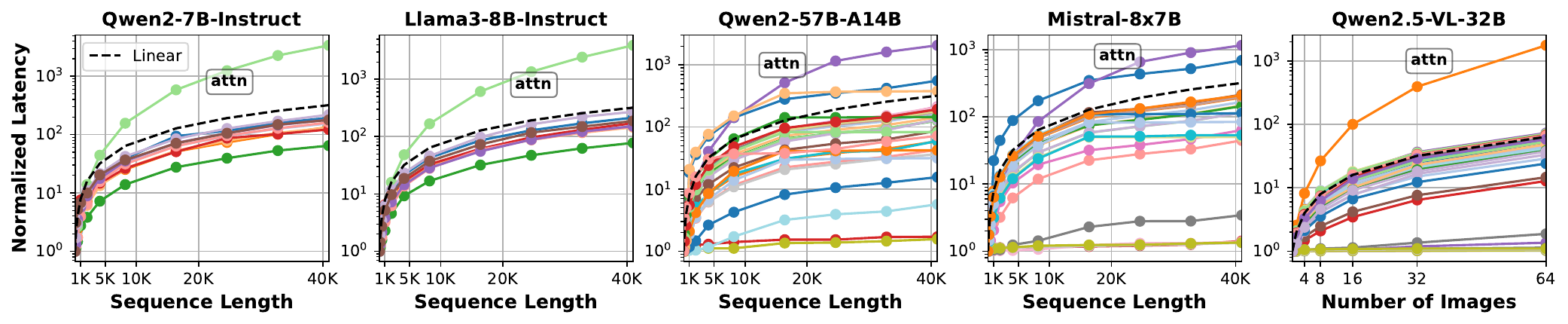}
    \caption{Compute sensitivity to input data sizes, for various operators in different model architectures.}
  \label{fig:comp-sensitivity}
\end{figure*}

%% file: figures/characterization/mem-sensitivity.tex
\begin{figure*}
    \centering
    \includegraphics[width=\textwidth]{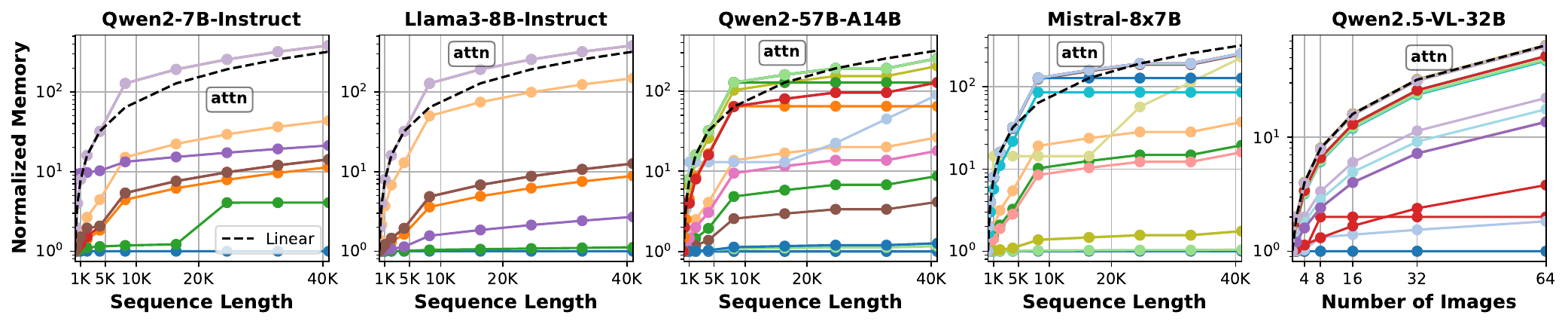}
    \caption{Memory sensitivity to input data size, for various operators in different model architectures.}
  \label{fig:mem-sensitivity}
\end{figure*}


%% file: figures/characterization/batch-size-sensitivity.tex
\begin{figure}[!t]
    \centering
    \includegraphics[width=\linewidth]{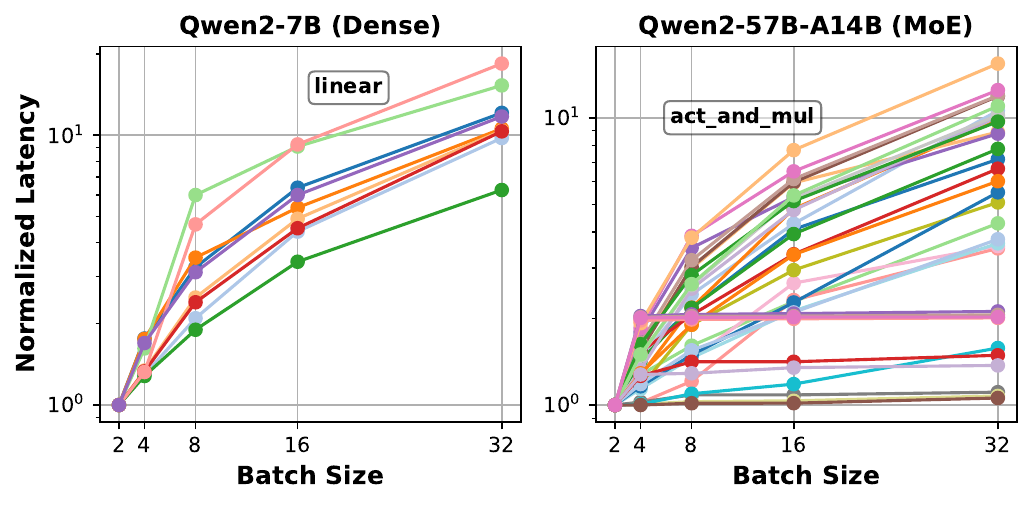}
    \caption{Compute sensitivity to batch sizes.}
    \label{fig:batching-sensitivity}
\end{figure}

%% file: figures/characterization/combined.tex
\begin{figure}[!t]
    \centering
    \includegraphics[width=\linewidth]{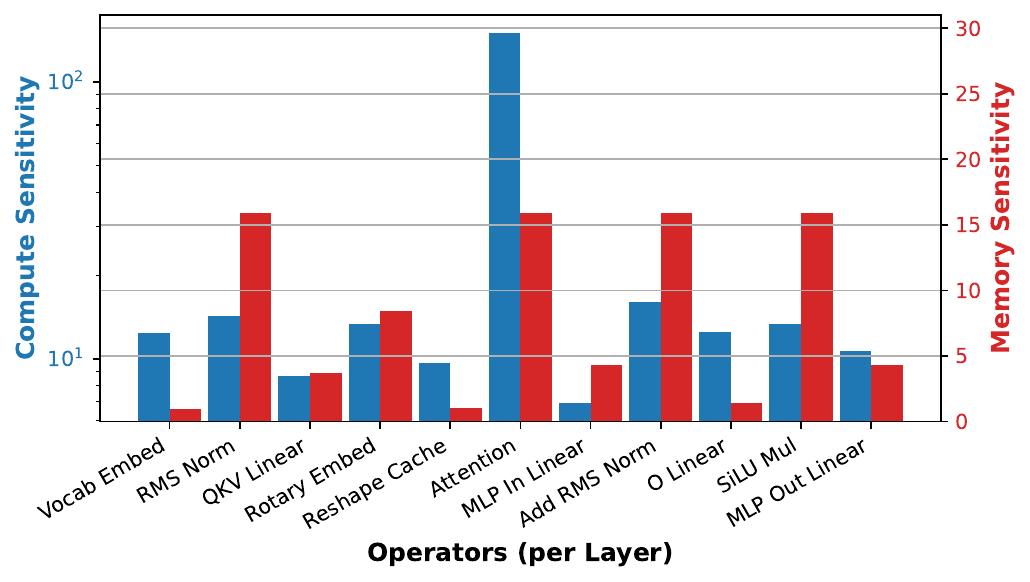}
    \caption{Compute- vs. memory-sensitive operators.}
    \label{fig:combined-comp-memory}
\end{figure}

%% file: figures/characterization/queueing-sensitivity.tex
\begin{figure}[!t]
    \centering
    \includegraphics[width=\linewidth]{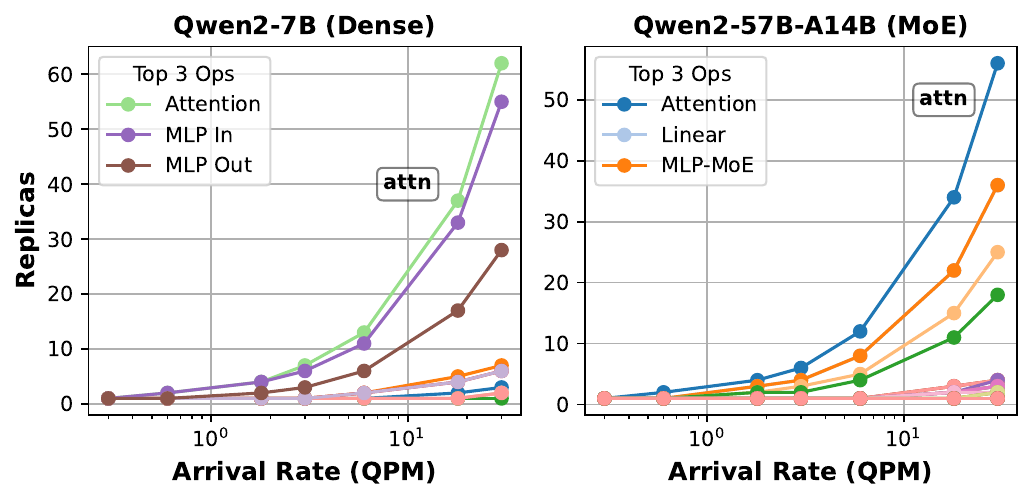}
    \caption{Queueing sensitivity to arrival rate.}
    \label{fig:queueing-sensitivity}
\end{figure}

%% file: figures/characterization/io-sensitivity.tex
\begin{figure}[!t]
    \centering
    \includegraphics[width=\linewidth]{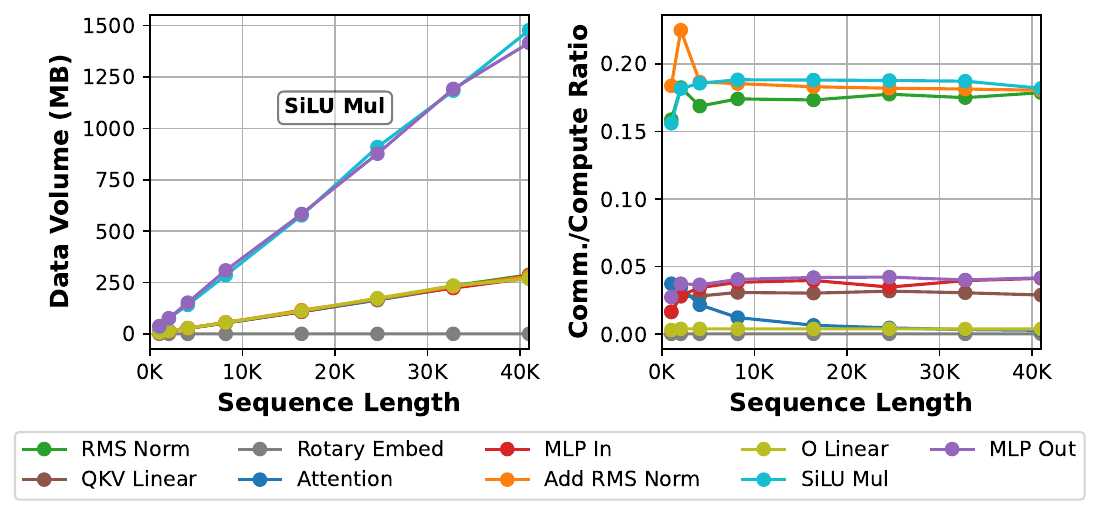}
    \caption{Operator input data volume for Qwen2-7B.}
    \label{fig:io-sensitivity}
\end{figure}

%% file: figures/characterization/sm-characterization.tex
\begin{figure}[!t]
  \centering
  \begin{subfigure}[b]{0.5\textwidth}
    \centering
    \includegraphics[width=1\textwidth]{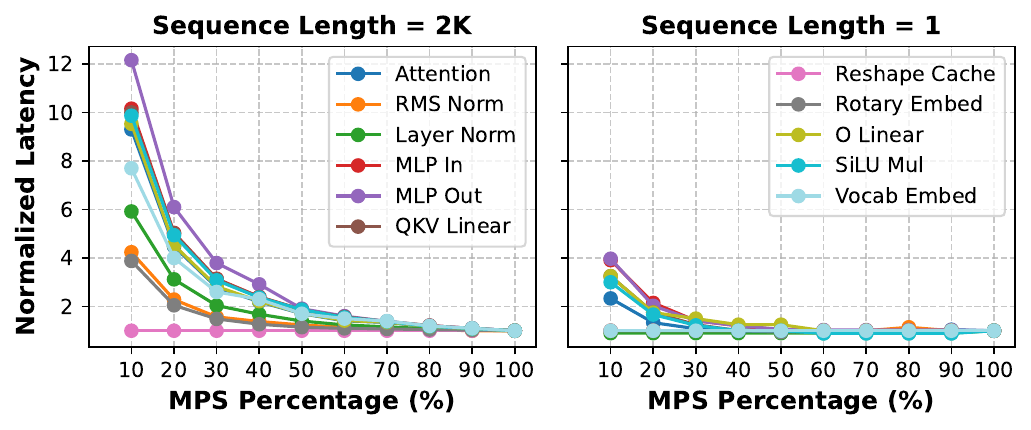}
    \caption{Operator normalized latency.}
    \label{fig:latency-mps}
  \end{subfigure}
  \begin{subfigure}[b]{0.5\textwidth}
    \centering
    \includegraphics[width=\textwidth]{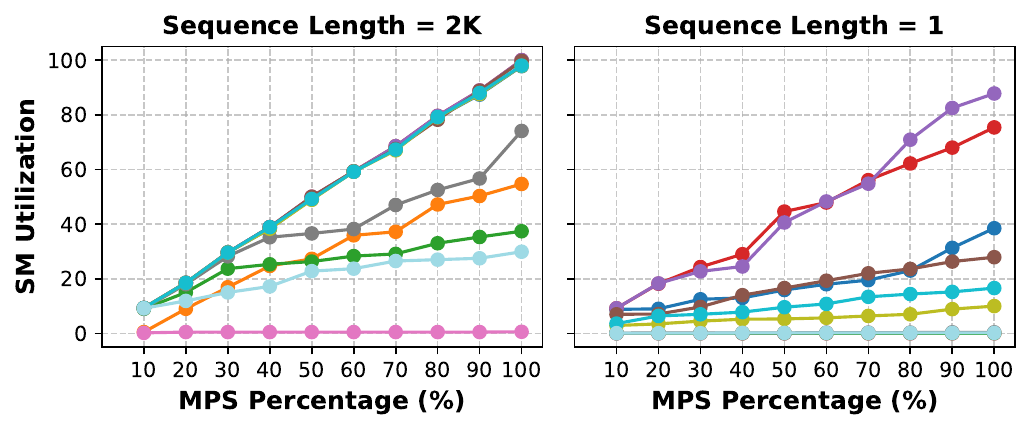}
    \caption{Operator SM utilization (\%).}
    \label{fig:util-mps}
  \end{subfigure}%
  \caption{Performance and SM utilization characteristics across operators (Qwen2~\cite{qwen2-7b}) under varying SM allocations.}
  \label{fig:sm-characterization}
\end{figure}

%% file: 003-analysis.tex
\section{A Theoretical Framework for Autoscaling}
\label{sec:analysis}

\begin{figure}[!t]
    \centering
    \includegraphics[width=0.95\linewidth]{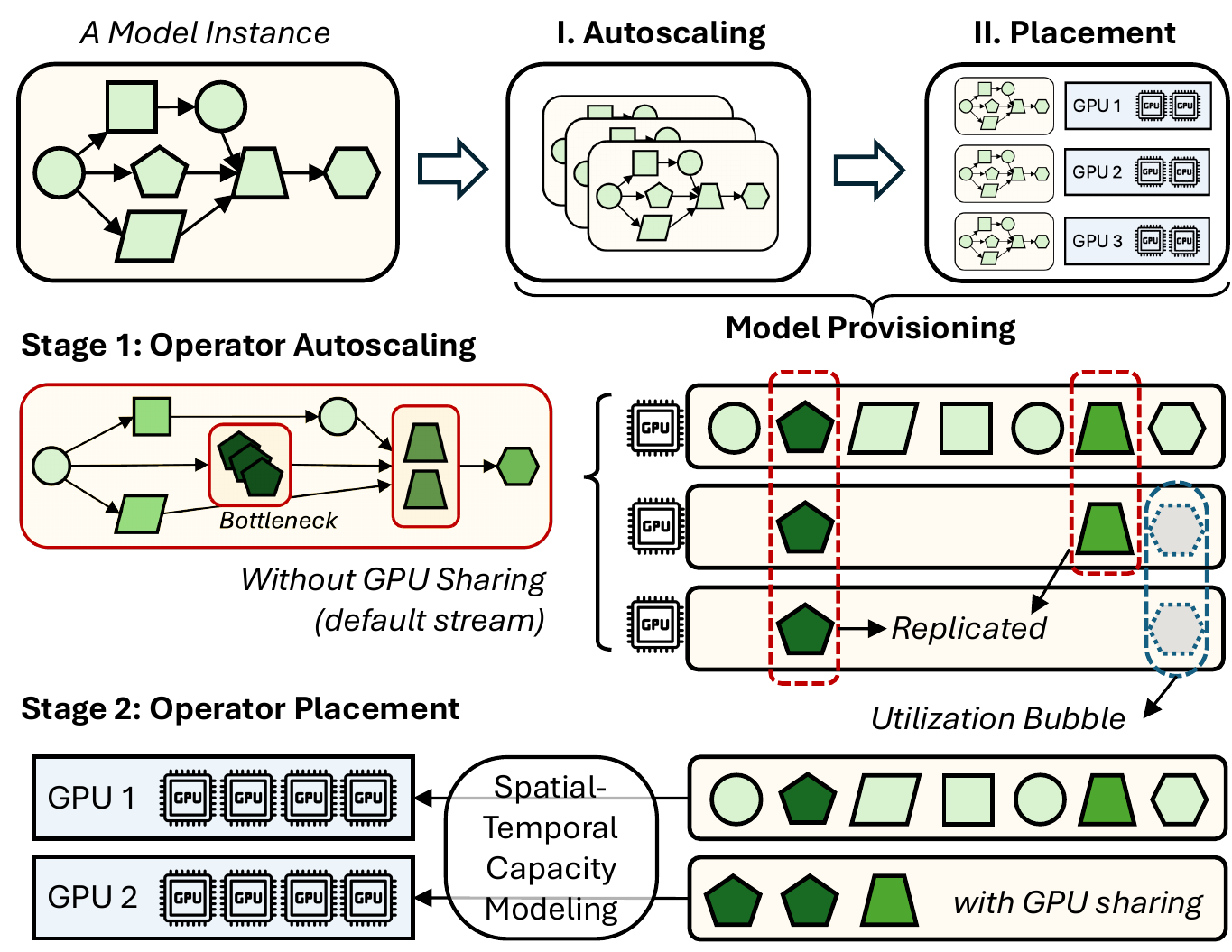}
    \caption{Model provisioning consists of (1) autoscaling and (2) placement at the operator granularity.}
    \label{fig:placement}
\end{figure}

Building on our systematic characterization and insights of operator performance and resource behavior, we now turn to a theoretical analysis of the benefits and the design of our operator-level model provisioning framework (\Cref{fig:placement}).
Specifically, we decompose the problem into two key stages: (1) operator autoscaling, and (2) operator placement, which mirrors autoscaling the whole model as the scaling unit and placing each model replica to fixed devices.

In this section, we first present a theoretical formulation for operator-level autoscaling in a model inference graph.
We use queueing theory to mathematically verify our insights and identify conditions broadly when operator-level autoscaling provides benefits.
At a high level, a computation graph consists of operators connected via data dependencies, with each operator characterized by computation time, memory consumption, and communication cost.
Given a stream of requests with a certain arrival rate (QPS) and request sequence (input) lengths, the goal for autoscaling is to scale each operator in terms of parallelism and replication to meet SLOs while minimizing total GPU usage.
In parallel, the goal for operator placement is to assign scaled operators to physical devices to minimize provisioning cost (\ie{}, devices) without SLO violations by modeling the spatial-temporal GPU utilization and capacity at the operator granularity.

\subsection{Problem Formulation}
\label{sec:analysis:formulation}

Consider a directed acyclic graph (DAG) of operators $\mathcal{G} = (\mathcal{V}, \mathcal{E})$, where $\mathcal{V}$ is the set of operators and $\mathcal{E}$ represents data dependencies. 
Let the input request stream $x \in X$ have an arrival rate $\lambda$ (requests per second) and a request sequence length distribution $L(x)$.
Each pass of the DAG corresponds to one \textit{iteration}.
For autoregressive models, a request goes through multiple iterations~\cite{patel2024splitwise}: (1) the first iteration is \textit{\textbf{prefill}}, which processes the full input sequence (\ie{}, length = input length); (2) subsequent \textit{\textbf{decode}} iterations (\ie{} length = 1) where each iteration generates an output token.

\myparagraph{Operator Attributes}
For each operator $v \in \mathcal{V}$:
\begin{itemize}[leftmargin=*,nosep]
    \item \textit{Computation time:} $T_{v} = f_v(P_v, L, B)$. The computation time of each operator depends on the model parallelism $P_v$, request sequence length $L$, and batch size $B$.
    \item \textit{Memory consumption:} $M_v = M^{\text{weight}}_v + M^{\text{transient}}_v$. Memory consumption of each operator consists of weight memory and transient memory (\eg{}, activation), which depends on the request sequence length $L$ and batch size $B$.
    \item \textit{Communication time:} $C_{v} = u_v(P_v, L, B)$ is the communication time of operator $v$ to its downstream operators, varying to parallelism $P_v$, sequence length $L$, and batch size $B$.
\end{itemize}

\myparagraph{Queueing Model}
Each operator is modeled as an $M/M/R_v$ queue, where $R_v$ is the number of replicas.
The service rate of each operator is referred to as $\mu_v = 1 / T_{v}$.
In queueing theory, the expected waiting time $W_v$ at the operator level is:
\begin{equation}
    W_{v} = \frac{C(R_v, \rho_v)}{R_v \mu_v - \lambda} \quad \text{with} \quad \rho_v = \frac{\lambda}{R_v \mu_v},
\end{equation}
where $C(R_v, \rho_v)$ is the Erlang-C formula:
\begin{equation}
    C(R_v, \rho_v) = \frac{(R_v \rho_v)^{R_v}}{R_v! \, (1 - \rho_v)} \Big/ \sum_{k=0}^{R_v-1} \frac{(R_v \rho_v)^k}{k!} + \frac{(R_v \rho_v)^{R_v}}{R_v! \, (1 - \rho_v)}.
\end{equation}

\myparagraph{Iteration Latency}
The latency for a single DAG iteration (\ie{}, one prefill iteration or one decode iteration) is
\begin{equation}
    T_{\text{total}} = \sum_{v \in \text{critical path}} (T_{v} + W_{v} + C_{v}).
\end{equation}

\myparagraph{Configurations}
Operator configurations are modeled as decision variables that include parallelism degree $P_v$, replicas $R_v$, and batch sizes $B_v$.
In addition, we consider the operator-to-device assignment $A_v$, which impacts both communication overhead and memory feasibility.
Another configuration dimension is the MPS share~\cite{mps}, which enables GPU sharing among operator replicas by specifying the fraction of allocated SM cores.
\begin{align}
    & P_v \in \mathcal{P}_v = \{1, 2, 4, 8, ...\}, \quad \forall v \in \mathcal{V} \\
    & R_v, B_v \in \mathbb{Z}^+, \quad \forall v \in \mathcal{V} \\
    & M_v \in \mathbb{Z}, \quad M_v \in [1, 100], \quad \forall v \in \mathcal{V}
\end{align}

\myparagraph{Constraints and Objectives}
TTFT and TBT SLOs on the inference latency are modeled as constraints, requiring that the prefill iteration latency is below TTFT SLO while the decode iteration latency meets TBT SLO, \ie{}, $T_{\text{total}} \leq T_{\text{SLO}}$.
In addition, for each device $d \in \mathcal{D}$, the aggregate memory consumption of all operators assigned to $d$ cannot exceed its memory capacity $M_d^{\text{cap}}$:
\begin{equation}
    \sum_{v \in \mathcal{V}: A_v = d} M_v \;\le\; M_d^{\text{cap}}, 
    \quad \forall d \in \mathcal{D}.
\end{equation}
Subject to this constraint, the objective is to minimize the aggregate GPU usage across all operators:
\begin{equation}
    \min \sum_{v \in \mathcal{V}} P_v \cdot R_v,
\end{equation}
where $P_v$ is the degree of parallelism and $R_v$ is the number of replicas for operator $v$.

\subsection{Resource Management Optimization}
\label{sec:analysis:optimization}

The optimization problem formulation (\Cref{sec:analysis:formulation}) determines the best configurations for each operator $v$ to minimize GPU usage while satisfying latency SLOs, taking into account computation, memory, communication, and queueing delays.
However, solving such an optimization problem exactly at the operator level is computationally expensive.
The configuration search space grows rapidly with the number of operators and their parameters (\eg{}, replicas, parallelism choices), making optimal solutions impractical to obtain at fine time granularity (\eg{}, every ten seconds).
To address this, we next present algorithms that approximate the optimal solution at significantly lower overhead by decoupling autoscaling from operator placement with device sharing.

\subsubsection{Operator-level Autoscaling}
At runtime, each operator's throughput and latency are primarily determined by how it is parallelized, replicated, and batched.
In the baseline model-level parallelism configuration~\cite{vllm-tp-pp}, tensor parallelism distributes model shards across devices within a server (constrained by memory capacity and model size), while pipeline parallelism connects stages across servers.
Each operator inherits this initial parallel structure from the model's deployment plan, which defines its starting parallelism degree $P_v$.
Building on this baseline, the operator autoscaler (Stage 1 in \Cref{fig:placement}) dynamically adjusts $P_v, R_v, B_v$ for each operator to (1) satisfy latency SLOs on TTFT or TBT, and (2) greedily minimize resource usage ($\sum_v P_v \cdot R_v$).
We refer to the full algorithm in \Cref{alg:greedy-autoscale}.

First, for each operator, we scan $b\in\{1,\ldots,B_v^{\max}\}$ with an initial $P_v$ inherited from model-level parallelism, set
$R_v(b) \leftarrow \big\lceil \lambda_v / \mu_v(b,p_v) \big\rceil$, and select the $(B_v,R_v)$ that minimizes the total sojourn time $\sum S_{v} = T_v(P_v,b) + W_v(\lambda_v,R_v(b),T_v)$ while maintaining stability~\cite{stability,harchol2013performance}.
This gives a set of low-latency, stable per-operator configurations that seed the global greedy search.

The global search proceeds iteratively. We repeatedly evaluate the current iteration latency $T_{total}$ along the critical path.
If $T_{total} > SLO$, we \emph{upscale} at the current bottleneck operator $v$ (the operator on the critical path with the largest $S_v$).
Upscaling prefers the smallest change that most effectively reduces $T_{total}$: increasing $R_v$ by one, optionally co-tuning $(B_v,P_v)$ to exploit batching or parallelism efficiency improvements.
After each candidate move, we recompute $\lambda$, update all affected sojourn times $S_v$, and re-evaluate $T_{total}$, accepting the move that maximally reduces $T_{total}$ (or minimally increases resource usage while restoring $T_{total}\le SLO$).

If $T_{total} < SLO$ (by more than a tolerance buffer $\varepsilon$), we attempt to release resources by downscaling the bottleneck operator $v$ on the critical path.
The candidate moves include: decrease $R_v$ by one (if stable) and optionally adjust $(B_v,P_v)$ to compensate.
Among feasible moves that keep $T_{total}\le SLO$, we select the one with the best objective (\eg{}, largest reduction in $\sum_v P_v \cdot R_v$ or total compute cost).
The loop terminates when no local move can improve the objective without violating the SLO (within $\varepsilon$), or when $T_{total}$ cannot be restored to the SLO via upscaling (\ie{}, infeasible SLOs).

The search space is discrete and finite: $R_v\in\mathbb{N}_+$, $B_v\in\{1,\ldots,B_v^{\max}\}$, $P_v\in\mathcal{P}_v$.
Each greedy step only considers local changes at the bottleneck, which concentrates optimization effort along the critical path where it most affects the iteration latency.
Since batch size influences both service rate and queuing, we always recompute arrival rates and sojourn times after every accepted move.
While not globally optimal, this greedy algorithm converges quickly to SLO-feasible, resource-efficient configurations and is lightweight enough to run online as workload traffic evolves.

\begin{algorithm}[!t]
\caption{Greedy Operator-Level Autoscaling}
\label{alg:greedy-autoscale}
\begin{algorithmic}[1]
\Require DAG $\mathcal{G}=(\mathcal{V},\mathcal{E})$, QPS $q$, batch limits $B_i^{\max}$, parallelism sets $\mathcal{P}_v$, latency functions $T_v(b,p)$ from profiling, SLO $T_{slo}$ with a buffer $\varepsilon$
\State Initialize $p_v \gets \min \mathcal{P}_v$ and $b_v \gets 1$ for all $v\in \mathcal{V}$
\ForAll{$v\in \mathcal{V}$} \Comment{Per-operator initialization}
    \State $\lambda \gets \textsc{ArrivalRates}(\mathcal{G}, q, b_v)$
    \State $r_v \gets \left\lceil \lambda / \mu_v(1,p_v) \right\rceil$ \quad where $\mu_v(b,p)=b/T_v(b,p)$
    \State $(b_v,r_v) \gets \arg\min\limits_{b \in \{1,\ldots,B_v^{\max}\}} \; s_v\!\left(\lambda,\left\lceil \lambda/\mu_v(b,p_v)\right\rceil,b,p_v\right)$
\EndFor
\State Recompute $\lambda \gets \textsc{ArrivalRates}(\mathcal{G}, q, \{b_v\})$
\State $s_v \gets W_v(\lambda_v,r_v,\mu_v(b_v,p_v)) + T_v(b_v,p_v)/b_v$ for all $v$
\State $T \gets \textsc{CriticalPathLatency}(G,\{s_v\})$
\While{true}
    \If{$T \le T_{slo}-\varepsilon$} \Comment{Scaling down}
        \State $j \gets \textsc{BottleneckOnCriticalPath}(G,\{s_v\})$
        \State $\mathcal{M} \gets \{(r_j-1,b_j,p_j)\} \cup \{(r_j-1,b,p_j)\mid b\in[b_j,B_j^{\max}]\} \cup \{(r_j-1,b,p)\mid b\in[b_j,B_j^{\max}],\, p\in\mathcal{P}_j\}$
        \State Filter $\mathcal{M}$ with stability check $\lambda_j < (r_j')\,\mu_j(b',p')$
        \State For each $m\in\mathcal{M}$: tentatively recompute $T'$
        \State Choose $m^\star \in \arg\min\{ \textsc{Cost}(\mathbf{r}',\mathbf{p}') \mid T' \le T_{slo} \}$
        \If{$m^\star$ exists} apply $m^\star$, set $T\gets T'$, continue
        \Else \, \textbf{break} \Comment{No further safe downscale}
    \EndIf
    \ElsIf{$T > T_{slo}$} \Comment{Scaling up}
        \State $j \gets \textsc{BottleneckOnCriticalPath}(G,\{s_v\})$
        \State $\mathcal{M} \gets \{(r_j+1,b_j,p_j)\} \cup \{(r_j+1,b,p_j)\mid b\in [1,B_j^{\max}]\} \cup \{(r_j+1,b,p)\mid b\in[1,B_j^{\max}],\, p\in\mathcal{P}_j\}$
        \State For each $m\in\mathcal{M}$: tentatively re-evaluate $T'$
        \State Choose $m^\star \in \arg\max\{T-T'\}$; Prefer the smallest $\Delta r_j$ that achieves $T' \le T_{slo}$)
        \If{$m^\star$ exists} apply $m^\star$, set $T\gets T'$, continue
        \Else \, \textbf{break} \Comment{Cannot improve further}
        \EndIf
    \Else
        \, \textbf{break} \Comment{Within tolerance of SLO}
    \EndIf
\EndWhile
\State \textbf{return} $\{(r_i,b_i,p_i)\}_{i\in V}$
\end{algorithmic}
\end{algorithm}

\subsubsection{Operator-to-Device Placement}

\Cref{alg:greedy-placement} presents the algorithm that maps operator replicas to devices while minimizing device usage under memory and SLO constraints.
We first compute the baseline $k_{\text{base}}=\min_v r_v$ and deploy those full model instances to form $\mathcal{D}_{\text{base}}$.
Extra replicas $\mathcal{R}_{\text{extra}}$ are sorted by $T_v$ (largest first) and placed greedily: for each $(v,k)$ we probe devices in $\mathcal{D}_{\text{base}}$, reject any that violate memory $M_d$ or where the interference-adjusted latency $T_v' = T_v \cdot I_{d,v}(b_v,p_v)$ would make the recomputed end-to-end latency exceed the SLO, and score feasible candidates by weighted residual slack (memory and compute) choosing the best; if none fit, a new device is provisioned.
As illustrated in \Cref{fig:placement} (Stage 2), the extra scaled-out replicas are preferentially colocated with existing model instances, and only provisioned on new devices when memory or compute capacity is insufficient.

\begin{algorithm}[!t]
\caption{Greedy Operator-Level Placement}
\label{alg:greedy-placement}
\begin{algorithmic}[1]
\Require DAG $\mathcal{G}=(\mathcal{V},\mathcal{E})$, config $\{(r_v,b_v,p_v)\}_{v\in\mathcal{V}}$ from Alg.~\ref{alg:greedy-autoscale}, device set $\mathcal{D}$, device capacities $\{M_d,U_d\}_{d\in\mathcal{D}}$, interference model $I_{d,v}(b,p)\ge 1$ (from profiling).
\State $k_{\text{base}} \leftarrow \min_{v\in\mathcal{V}} r_v$ \Comment{Number of full model instances}
\State Construct replica sets:
\State \quad $\mathcal{R}_{\text{base}} \leftarrow \{(v,i)\mid v\in\mathcal{V},\; i\in[1,k_{\text{base}}]\}$
\State \quad $\mathcal{R}_{\text{extra}} \leftarrow \{(v,i)\mid v\in\mathcal{V},\; i\in[k_{\text{base}}+1,\,r_v]\}$
\State Sort $(v,k) \in \mathcal{R_{\text{extra}}}$ in descending order of $T_v$
\State $\mathcal{D}_{\text{base}} \leftarrow \textsc{DeployModelInstance}(\mathcal{R}_{\text{base}})$
\State $\mathcal{D}_{\text{empty}} \gets \mathcal{D} \setminus \mathcal{D}_{\text{base}}$
\ForAll{$(v,k)\in\mathcal{R}_{\text{extra}}$}
    \State $Candidates \gets \emptyset$
    \ForAll{$d\in\mathcal{D}_{\text{base}}$} \Comment{Try existing devices first}
        \If{$MemLoad_d + m_v > M_d$} \State \textbf{continue} \EndIf
        \State $T_v' \gets T_v\cdot I_{d,v}(b_v,p_v)$
        \If{$\textsc{ReComputeLatency}(\mathcal{G}) > SLO$} \State \textbf{continue} \EndIf
        \State $slack\_mem \gets M_d - (MemLoad_d + m_v)$
        \State $slack\_comp \gets U_d - (CompLoad_d + T_v')$
        \State $Candidates.\textsc{Append}(d)$
    \EndFor
    \If{$Candidates == \emptyset$} \Comment{No existing device fits}
        \State $d_{\textsf{new}} \gets \textsc{ProvisionDevice}(\mathcal{D}_{\text{empty}}, \mathcal{D}_{\text{base}})$
        \State $d^\star \gets d_{\textsf{new}}$
    \Else
        \State $\textsc{ComputeWeightedSlack}(Candidates)$
        \State $d^\star \gets \arg\max_{(d)\in Candidates} slack$
    \EndIf
    \State $Placement \gets \textsc{AssignOperator}(v,k,d^\star)$
\EndFor
\State \textbf{return} $Placement$
\end{algorithmic}
\end{algorithm}

\myparagraph{Default Stream Constraint}
To this end, we assume multi-stream is enabled on GPUs that support time or space sharing.
However, in older GPUs that do not support GPU sharing, replicas of the same operators are executed sequentially as a single stream in the default stream setup.
As a result, scaled-out replicas are placed on separate devices, eliminating the possibility of reducing the number of active GPUs compared to model-level autoscaling.
While this constraint removes opportunities for device savings, it opens the door for energy optimizations: lower compute density on each scaled-out device can reduce overall energy consumption.
To capture this effect, we introduce an operator-level energy attribution model for each request:
\begin{equation}
    E_v = \alpha_v \cdot P_v \cdot R_v \cdot (W_v + T_v) + \beta_v \cdot T_{v},
\end{equation}
where $\alpha_v$ and $\beta_v$ are power coefficients for device usage (idle power) and active computation (dynamic power).

\subsubsection{Baselines}

\myparagraph{Model-level Autoscaling and Provisioning}
As a comparison, model-level autoscaling treats the entire model as a monolithic unit, enforcing uniformity across all operators.
Specifically, all operators share the same batch size $B$ and the same number of replicas $R$, with parallelism $P$ fixed by the chosen tensor/pipeline partitioning strategy.
Rather than tuning per-operator parameters, the autoscaler adjusts $(B,R)$ globally to meet latency SLOs on TTFT or TBT.
Model-level autoscaling therefore provides a coarse-grained but stable baseline: it captures system-wide scaling trends but lacks the flexibility to exploit per-operator heterogeneity in workload intensity or compute efficiency, limiting opportunities for fine-grained resource optimization compared to operator-level autoscaling.
Every scaled-out model replica is placed onto a new set of GPU devices without sharing.

\myparagraph{Brute-force Approach}
As an oracle baseline, brute-force search enumerates all operator configurations $(P_v, R_v, B_v)$, evaluates end-to-end latency, and selects the resource-minimal SLO-feasible point.
While this guarantees optimality, the combinatorial space $O\left(\prod_v |\mathcal{P}_v|\cdot B_v^{\max}\cdot R_v^{\max}\right)$ makes it computationally prohibitive for online use.

\subsection{Experimental Analysis}
\label{sec:analysis:results}

We now evaluate the effectiveness of operator-level autoscaling and placement compared to conventional model-level provisioning.
Our goal is to quantify the resource and efficiency gains (\ie{}, in device usage, energy consumption, and memory utilization) while preserving latency SLOs.
Using the theoretical formulation and algorithms described in the previous section, we simulate autoscaling behavior without executing full model inference~\cite{lin2024nnscaler}.
We analyze how operator-level scaling adapts under varying workload and SLO conditions, including changes in sequence length, batch size, and request arrival rate (QPS).
Experiments are conducted using representative models, Qwen2-7B~\cite{qwen2-7b} and Qwen2-MoE~\cite{qwenmoe}, which capture both dense and sparse (MoE) inference characteristics.
To simulate real-world workloads, we adopt production LLM inference traces from Azure~\cite{stojkovic2025dynamollm} and Moonshot AI~\cite{qin2024mooncake}.
Through these experiments, we highlight when and why fine-grained operator-level autoscaling yields substantial savings over traditional model-level scaling, especially under heterogeneous or dynamic workloads.

\input{figures/analysis/savings-vs-seqlen}

\myparagraph{Varying Request Sequence Length}
We first examine how varying the input sequence length affects autoscaling behavior and resource allocation under SLOs, capturing the distinct scaling patterns between prefill- and decode-dominated workloads.
As shown in \Cref{fig:analysis-seqlen}, increasing sequence length yields varying savings across GPU devices and energy, and memory resources, with diminishing returns beyond 8K tokens.
GPU savings (\Cref{fig:analysis-seqlen-gpus}) peak around 30\% at 4K tokens for the dense model and 40\% for MoE as operator-level provisioning effectively consolidates workloads during longer prefill phases, but drop at very long sequences where SM saturation limits further gains from GPU sharing.
MoE models benefit from higher savings because of their sparsity and more diversity in operator sensitivity.
Energy savings (\Cref{fig:analysis-seqlen-energy}) follow a similar trend, reaching up to 25\% reduction at peak, mainly due to the savings in devices.
When there is no savings in the GPU devices, the energy consumption for operator-level provisioning is low due to only scaling out bottlenecked operators.
Memory savings (\Cref{fig:analysis-seqlen-memory}) grow more steadily with sequence length, surpassing 60\% at 32K tokens for Qwen2-7B and 64K tokens for Qwen-MoE.
Unlike GPU and energy savings, this trend arises not from operator colocation but from selective scaling: only compute-bottlenecked operators are scaled out, yielding substantial memory reductions from the non-sensitive operators, compared to uniformly scaling all operators in model-level provisioning.

\boxinsight{Operator-level autoscaling yields the largest benefits at moderate sequence lengths, before SM contention tightens SLO margins and reduces flexibility at very long contexts. Even so, it achieves up to 25\% energy and 60\% memory savings across workloads.}
\label{insight:seqlen-analysis}

\input{figures/analysis/analysis-vs-qps}

\myparagraph{Varying QPS}
Next, we analyze the impact of request arrival rate (QPS) on operator-level versus model-level autoscaling, highlighting how operator-level granularity enables more resource-efficient scaling decisions to mitigate queueing delays.
As shown in \Cref{fig:analysis-qps}, operator-level autoscaling consistently delivers higher resource savings across GPU, energy, and memory compared to model-level provisioning, especially at moderate QPS.
For a dense model Qwen2-7B, GPU savings (\Cref{fig:analysis-qps-gpus}) peak around 30\% near 40 QPS, where operator-level scaling consolidates workloads and avoids over-provisioning during bursts.
Beyond this point, savings fluctuate slightly as SM saturation and tighter latency margins reduce opportunities for GPU sharing.
Energy savings (\Cref{fig:analysis-qps-energy}) exhibit a similar pattern, reaching up to 25\% at high QPS for Qwen2-7B, primarily driven by reduced device counts and selective operator scaling rather than uniform expansion.
Memory savings (\Cref{fig:analysis-qps-memory}) grow steadily with QPS, surpassing 50\% at 100 QPS, reflecting a similar trend with sequence length growth.

The Qwen-MoE models exhibit a similar scaling behavior in terms of resource savings.
For both models, GPU savings remain negligible at low QPS (<20), where the provisioned model instance can handle limited traffic demands without scaling under the SLOs.

\boxinsight{Operator-level autoscaling achieves its largest benefits under moderate to high QPS, where fine-grained scaling mitigates queueing delays without excessive resource inflation, resembling the sequence length trend.}
\label{insight:qps-analysis}

\input{figures/analysis/analysis-vs-phases}

\myparagraph{Prefill vs. Decode}
We compare operator-level and model-level provisioning behavior across the prefill (known to be compute-bound~\cite{patel2024splitwise}) and decode stages (memory-bound), illustrating how their distinct computational and temporal characteristics drive different scaling needs.
We analyze two production-scale LLM inference traces from (1) Azure LLM inference cluster~\cite{stojkovic2025dynamollm} and (2) Mooncake LLM serving platform~\cite{qin2024mooncake}.
As shown in \Cref{fig:analysis-stage}, operator-level resource management yields consistently higher savings during the prefill stage across all workloads.
On Azure traces, prefill achieves up to 35\% GPU, 25\% energy, and 45\% memory savings for chat services, while decode savings remain modest (below 15\%).
Savings in coding services are less due to its low QPS in the traces.
Similarly, in Mooncake traces, prefill achieves up to 22\% GPU savings, 20\% energy, and 41\% memory, higher than the savings achieved during decode.
These results highlight that prefill stages benefit more from operator-level provisioning and scaling due to its denser compute utilization and shorter execution bursts.

\boxinsight{Prefill stages offer substantially higher optimization potential than decode—up to 2--3\texttimes{} greater resource savings, as they are more compute-intensive and bursty, making them ideal targets for fine-grained operator-level model provisioning and scaling.}

\myparagraph{Large vs. Small Models}
We analyze how model size influences autoscaling effectiveness using the Qwen2 family, assuming all models are served under the same SLO target to isolate the effect of scale.
As shown in \Cref{fig:analysis-layers}, operator-level provisioning yields substantial gains across all model sizes, with increasing benefits as model scale grows.
For smaller models like Qwen2-0.5B, savings are modest (below 15\%) since coarse-grained model-level provisioning is already sufficient to meet SLOs.
As model size increases to Qwen2-1.5B and Qwen2-7B, GPU and energy savings rise to around 20–30\%, while memory savings exceed 35\%, driven by more diverse operator-level utilization patterns that can be effectively co-scheduled.
For the largest model, Qwen2-72B, both energy and memory savings peak near 50\%, highlighting that large models benefit most from operator-level provisioning, which dynamically reallocates resources to mitigate fragmentation and idle GPU spaces.
If the SLO were instead proportional to model size, we would expect the relative savings to remain similar across models, as larger models effectively replicate the same layer structure, leading to comparable operator-level efficiency patterns.

\boxinsight{Larger models amplify the benefits of operator-level autoscaling under a fixed SLO, while proportional SLO scaling yields similar relative savings since operator-level scaling patterns remain consistent across sizes.}
\label{insight:model-size}

\input{figures/analysis/analysis-vs-layers}

\noindent
\textbf{How Far from the Oracle?}
We quantify the optimality gap between the proposed operator-level autoscaling algorithm (\Cref{sec:analysis:optimization}) and the brute-force oracle.
Across workload conditions with varying QPS at a sequence length of 1K, the average resource savings gap is 8\%, suggesting that the greedy algorithm attains most of the theoretical optimum with significantly lower computational overhead.

%% file: figures/analysis/savings-vs-seqlen.tex
\begin{figure}[!t]
  \centering
  \begin{subfigure}[b]{0.5\textwidth}
    \centering
    \includegraphics[width=1\textwidth]{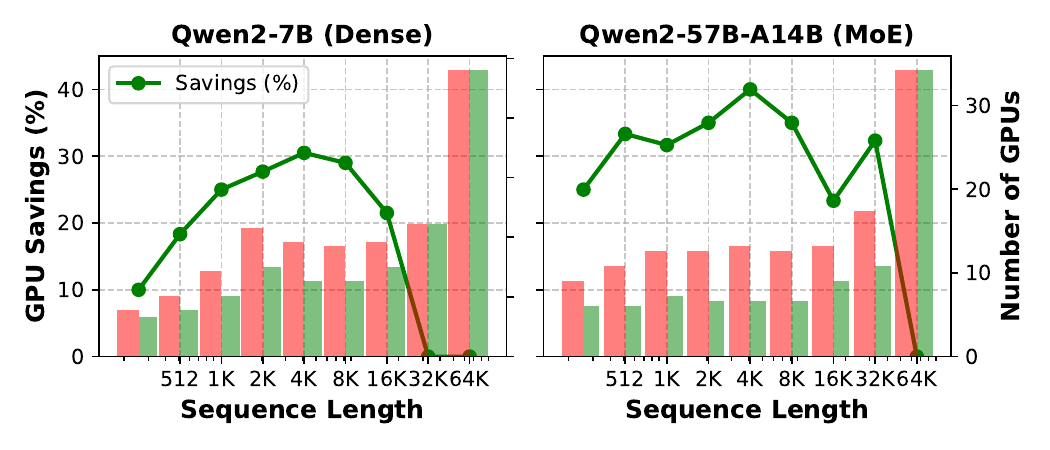}
    \caption{GPU savings.}
    \label{fig:analysis-seqlen-gpus}
  \end{subfigure}
  \begin{subfigure}[b]{0.5\textwidth}
    \centering
    \includegraphics[width=\textwidth]{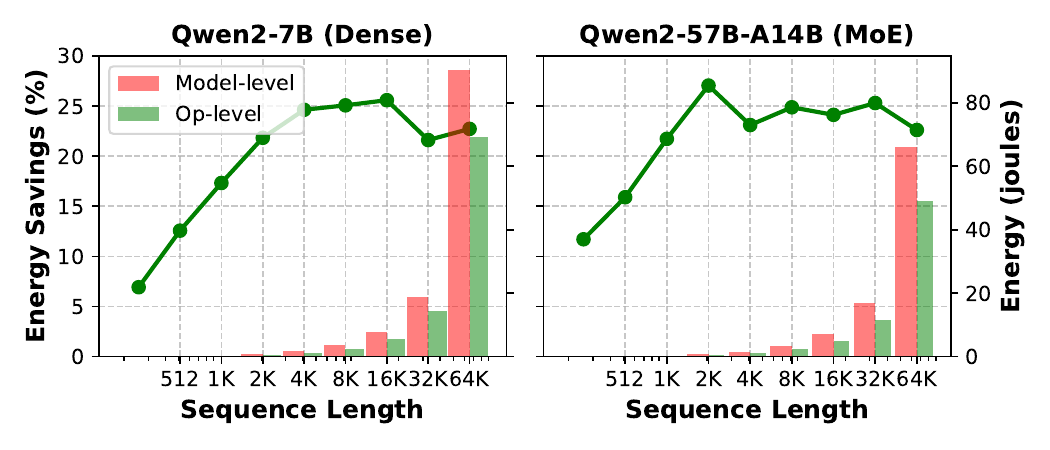}
    \caption{Energy savings.}
    \label{fig:analysis-seqlen-energy}
  \end{subfigure}
  \begin{subfigure}[b]{0.5\textwidth}
    \centering
    \includegraphics[width=\textwidth]{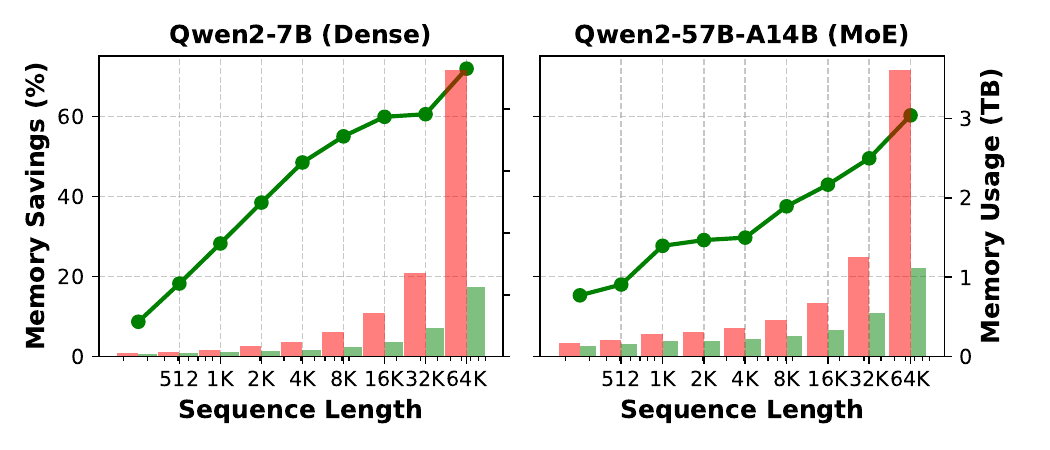}
    \caption{Memory savings.}
    \label{fig:analysis-seqlen-memory}
  \end{subfigure}%
  \caption{Benefits of operator-level resource management under varying sequence lengths.}
  \label{fig:analysis-seqlen}
\end{figure}

%% file: figures/analysis/analysis-vs-qps.tex
\begin{figure}[!t]
  \centering
  \begin{subfigure}[b]{0.5\textwidth}
    \centering
    \includegraphics[width=1\textwidth]{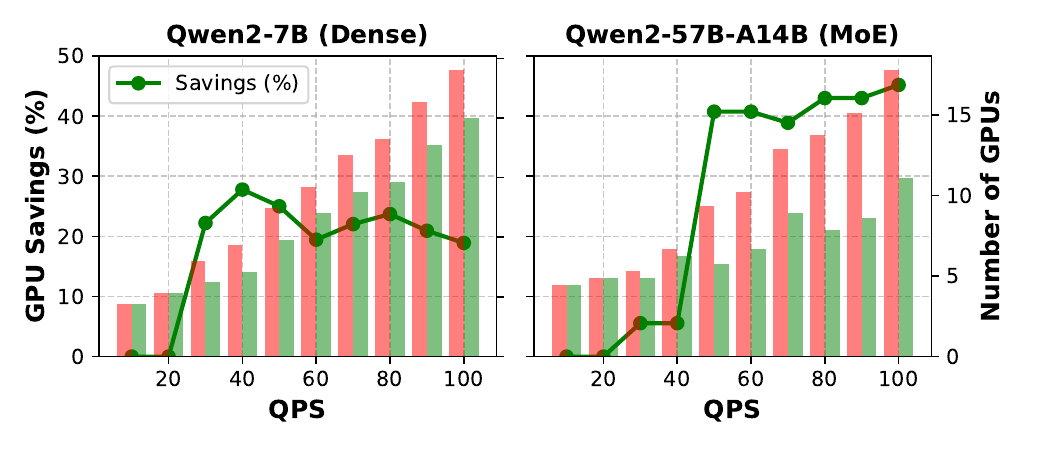}
    \caption{GPU savings.}
    \label{fig:analysis-qps-gpus}
  \end{subfigure}
  \begin{subfigure}[b]{0.5\textwidth}
    \centering
    \includegraphics[width=\textwidth]{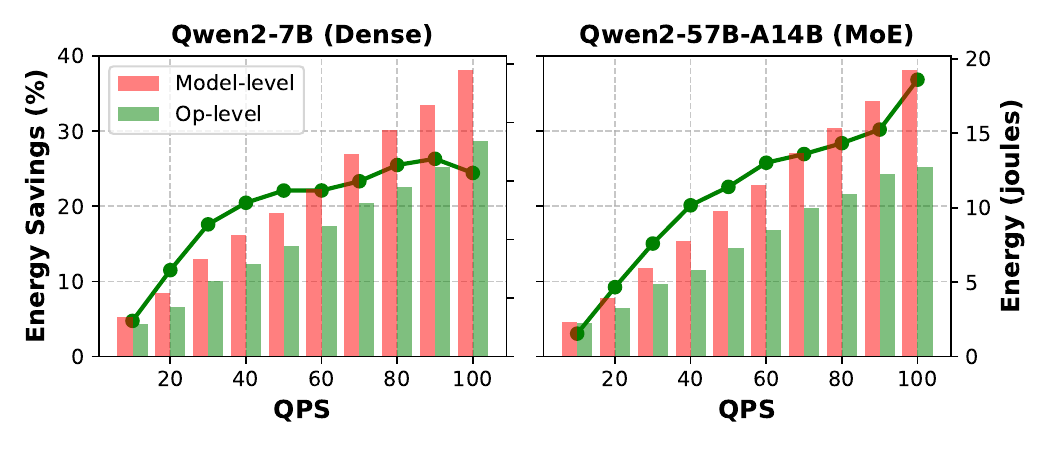}
    \caption{Energy savings.}
    \label{fig:analysis-qps-energy}
  \end{subfigure}
  \begin{subfigure}[b]{0.5\textwidth}
    \centering
    \includegraphics[width=\textwidth]{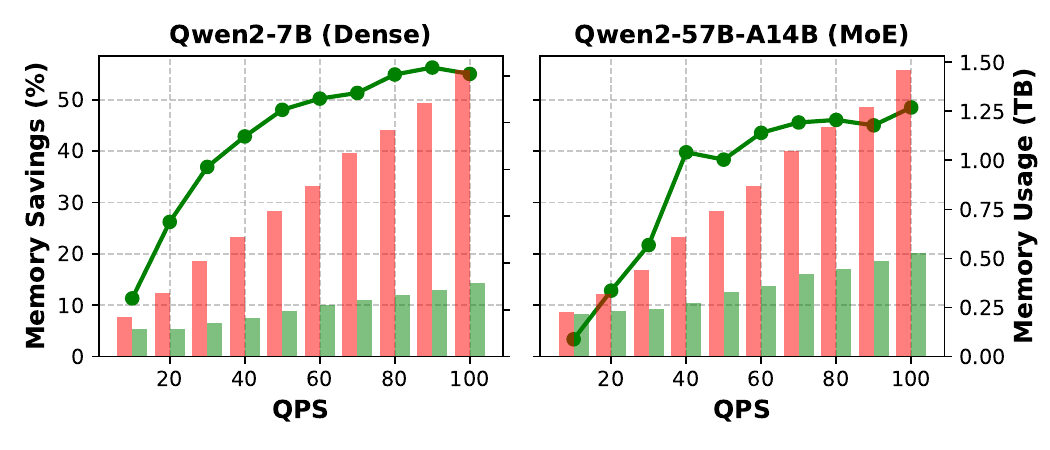}
    \caption{Memory savings.}
    \label{fig:analysis-qps-memory}
  \end{subfigure}%
  \caption{Benefits of operator-level resource management under varying QPS.}
  \label{fig:analysis-qps}
\end{figure}

%% file: figures/analysis/analysis-vs-phases.tex
\begin{figure}[!t]
    \centering
    \includegraphics[width=\linewidth]{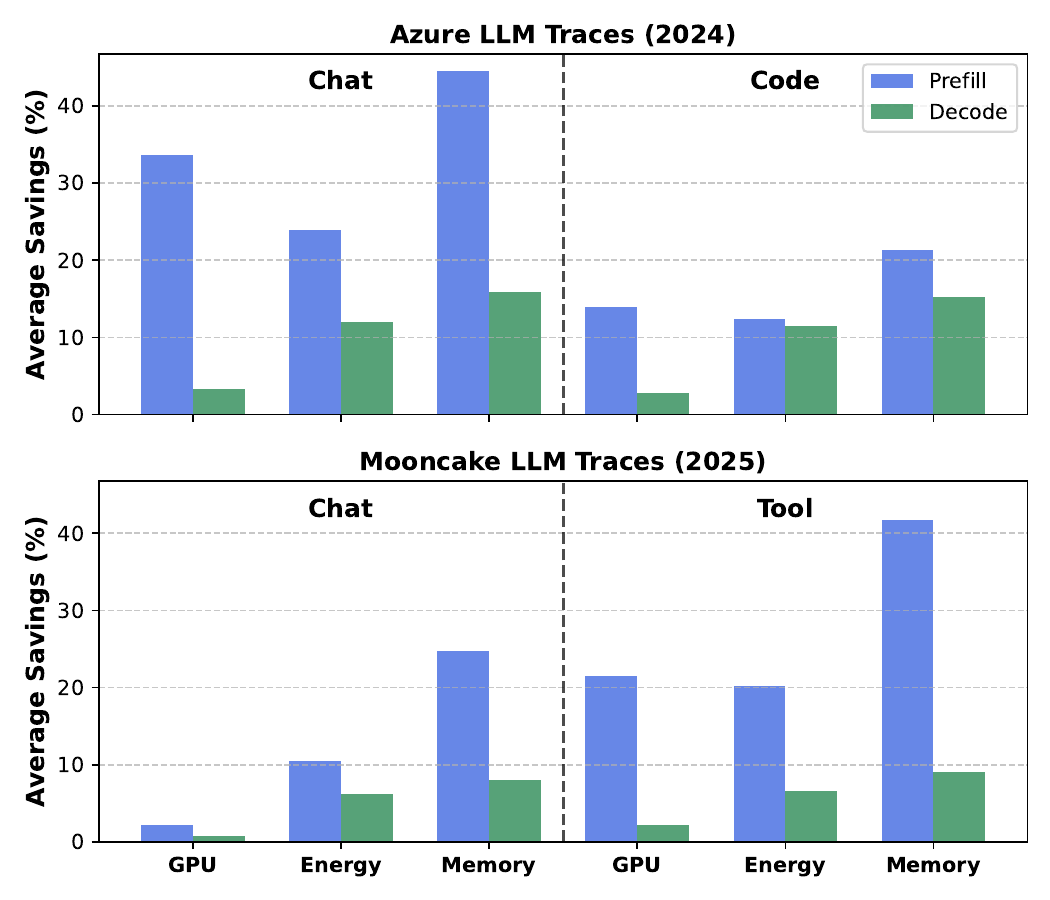}
    \caption{Benefits of operator-level resource management in prefill vs. decode stages for Qwen2-7B.}
    \label{fig:analysis-stage}
\end{figure}

%% file: figures/analysis/analysis-vs-layers.tex
\begin{figure}[!t]
    \centering
    \includegraphics[width=\linewidth]{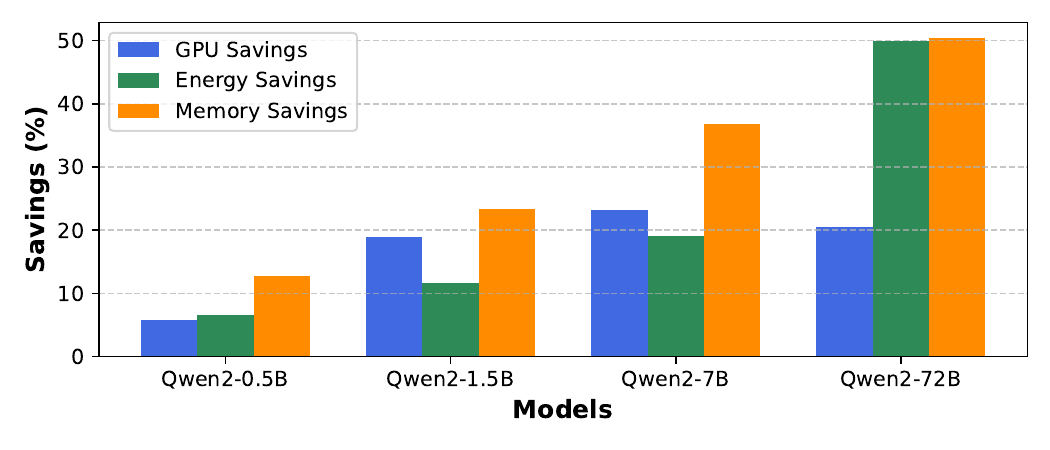}
    \caption{Benefits of operator-level resource management across various-sized models}
    \label{fig:analysis-layers}
\end{figure}

%% file: 006-discussion.tex
\section{Discussion}

\myparagraphnodot{When to Adopt Operator-level Autoscaling?}
Operator-level autoscaling is most beneficial when workloads are latency-tolerant and dominated by heterogeneous operator behaviors.
In contrast, inference systems pursuing ultra-low latency through megakernel fusion~\cite{megakernel} (where all operators are fused into a single large kernel) leave limited room for operator-level provisioning and scaling.
Thus, operator-level autoscaling strikes a balance between efficiency and flexibility, favoring modular deployments and scaling over monolithic kernels.

\myparagraphnodot{Prefill-Decode Disaggregation or Co-location?}
Our analysis (\Cref{sec:analysis:results}) shows that prefill stages, being compute-intensive and variable in demand, gain the most from fine-grained autoscaling.
Decode, while often constrained by tighter latency budgets, benefits less from operator-level provisioning and scaling.
This suggests that enabling prefill–decode disaggregation complements operator-level scaling by exposing varying extents of optimization opportunities across stages.
In prefill-decode co-location cases with chunked prefill, the maximum sequence length is effectively bounded by the chunk size, altering operator load distribution.
Operator-level autoscaling remains valuable in such cases, as it can adapt provisioning to dynamic chunking~\cite{agrawal2024medha}.

\myparagraph{Integration with Inference Schedulers}
Operator provisioning and scaling should be co-optimized with request scheduling at the instance and cluster level to minimize inter-operator communication and resource fragmentation.
Integrating scaling logic with LLM inference schedulers allows coordinated resource allocation that respects both topology and operator affinity, which we leave to future work.


%% file: 007-related.tex
\section{Related Work}

\myparagraph{LLM Serving Optimization}
Prior work on optimized LLM serving, including batch scheduling~\cite{orca,aiops2024qiu,patke2024queue,sun2024llumnix}, chunked prefill~\cite{sarathi-serve}, KV-cache management~\cite{vllm,qin2024mooncake,hu2024memserve}, and energy or power management~\cite{polca,stojkovic2025tapas,stojkovic2025dynamollm}, are complementary to our contribution, as they target the serving engine and scheduler layers, whereas we focus on the \textit{model execution phase} through operator-level autoscaling and placement.

In addition, there has been a trend of disaggregation on generative model serving:
(1) Prefill-Decode disaggregation~\cite{patel2024splitwise,zhong2024distserve} separates the two phases of generation, enabling independent autoscaling.
(2) Encoder–LLM disaggregation~\cite{qiu2025modserve,epd,dong2025hydrainfer} decouples multimodal encoders from LLM backends, allowing modality-specific scaling.
(3) VAE-DiT disaggregation~\cite{huang2025ddit} enables specialized autoscaling in video/image generation pipelines.
(4) MA parallelism~\cite{wang2025step,zuo2025serving,zhu2025megascale,chowdhery2023palm,liu2025expert} disaggregates attention and MLP/MoE layers for independent deployment to achieve higher throughput.
From this perspective, our work pushes disaggregation to a finer granularity: the \textit{operator level}. We examine operator heterogeneity and the benefits of conceptually decoupling operators to enable fine-grained autoscaling and SM-aware placement, without necessarily disaggregating them across separate devices.

\myparagraph{Autoscaling Policies}
As mentioned in \Cref{sec:bg:autoscaling}, AIBrix~\cite{team2025aibrix}, DynamoLLM~\cite{stojkovic2025dynamollm}, Chiron~\cite{patke2025hierarchical}, and DeepServe~\cite{deepserve} propose model-level autoscaling \textit{policies} that combine demand prediction with replica management.
Our work is orthogonal: rather than policy design, we contribute a new autoscaling \textit{mechanism} at the operator level, which benefit from advanced autoscaling policies in this field.

\myparagraph{Enabling Multi-Stream for Efficiency}
Recent work~\cite{lin2025bullet,zhu2025nanoflow,kamath2025pod} has leveraged multi-stream processing on GPUs to improve utilization and throughput.
For example, NanoFlow~\cite{zhu2025nanoflow} exploits intra-device parallelism by overlapping computation with I/O, thereby increasing LLM serving throughput.
Similarly, Pod-Attention~\cite{kamath2025pod} collocates prefill and decode stages to jointly utilize compute and memory bandwidth.

\myparagraph{Operator-level Optimization}
Several efforts have explored operator-level optimizations from complementary perspectives, distinct from our focus on autoscaling and fine-grained resource management.
For instance, $\mu$-Serve~\cite{qiu2024power} applies operator-level GPU frequency scaling for power efficiency; MegaKernel~\cite{wu2024mirage} fuses all operators into a single kernel to minimize latency; operator fusion techniques~\cite{li2024large} reduce launch overhead; and customized attention kernels such as FlashInfer~\cite{ye2025flashinfer} optimize critical operators for lower latency.

\myparagraph{Multi-model Multiplexing}
In GPU clusters that serve multiple LLMs, prior work has explored spatial and temporal multiplexing to improve cluster utilization and reduce serving costs~\cite{duan2024muxserve,li2023alpaserve,xiang2025aegaeon,xing2025towards,yu2025prism,qiao2025conservefinegrainedgpuharvesting}.
While this paper focuses on single-model serving, its design naturally extends to multi-model scenarios. By independently managing the scaling and provisioning of each model, it enables faster switching across models and more flexible spatial sharing of limited GPU resources.

%% file: 008-conclusion.tex
\section{Conclusion}

In this paper, we present a systematic characterization and theoretical framework for operator-level autoscaling and placement in large-model inference systems.
Through queueing-based modeling, we decompose inference provisioning into fine-grained operator units, each governed by computation, memory, and communication tradeoffs.
Our analysis demonstrates that, compared to traditional model-level scaling, operator-level resource management substantially improves GPU, memory, and energy efficiency under varying sequence lengths, request arrival rates, and model sizes—achieving up to 40\% GPU and 35\% energy savings while preserving SLOs.
These results highlight the potential of shifting from coarse, model-centric scaling to a fine-grained, \textbf{\textit{operator-centric inference architecture}}.

Looking ahead, our findings point to a future inference system design that exploits operator-level provisioning in the model runtime layer.
With continuous workload monitoring or prediction, heterogeneous operators can be co-scheduled across shared devices, and exploit spatial-temporal utilization patterns to minimize cost and energy while preserving latency SLOs.
This fine-grained, operator-aware resource management offers a foundation for the next generation of elastic and efficient large-scale inference infrastructures.